# X-ray imaging detector for radiological applications in the harsh environments of low-income countries

M.A. Chavarria[1], M. Huser[2], S. Blanc[1], P. Monnin[3], J. Schmid[4], C. Chênes[4], L. Assassi[4], H. Blanchard[5], R. Sahli[5], J. Thiran[6], R.P. Salathé[7] and K. Schönenberger[1*]

[1] *EssentialTech Centre, Ecole Polytechnique Fédérale de Lausanne (EPFL), Lausanne CH-1015, Switzerland*

[2] *Ecole technique - Ecole des métiers - Lausanne (ETML), Lausanne CH-1004, Switzerland*

[3] *Institute of radiation physics (IRA), Centre hospitalier universitaire vaudois (CHUV), Lausanne CH-1007, Switzerland*

[4] *Dept. of Medical radiology technology, Geneva School of Health Sciences, HES-SO, Genève CH-1206, Switzerland*

[5] *Pristem S.A. Lausanne CH-1007, Switzerland*

[6] *Signal Processing Laboratory 5, Ecole Polytechnique Fédérale de Lausanne (EPFL), Lausanne CH-1015, Switzerland*

[7] *School of Engineering, Ecole Polytechnique Fédérale de Lausanne (EPFL), Lausanne CH-1015, Switzerland*

\* Corresponding author

**E-mail:** klaus.schonenberger@epfl.ch (KS)

# Abstract

This paper describes the development of a novel medical X-ray imaging system adapted to the needs and constraints of low and middle-income countries. The developed system is based on an indirect conversion chain: a scintillator plate produces visible light when excited by the X-rays, then a calibrated multi-camera architecture converts the visible light from the scintillator into a set of digital images. The partial images are then unwarped, enhanced and stitched through parallel processing units (FPGA) and a specialized software. All the detector components were carefully selected focusing on optimizing the system's image quality, robustness, cost-effectiveness and capability to work in harsh tropical environments. With this aim, different customized and commercial components were characterized. The resulting detector can generate high quality medical diagnostic images with DQE levels up to 60 % (@ 2.34 µGy), even under harsh environments i.e. 60 ºC and 98% humidity.

# Introduction

Over the last decade, digitalization has played a prominent role in the X-ray medical equipment. Film-based and Computed Radiography (CR) plates are been replaced by fully digital X-ray detectors, commonly flat panels, requiring lower amount of dose and providing better image quality in a very short amount of time. In addition, digital technology enables video capabilities, which has given rise to new applications, such as fluoroscopy, Cone Beam Computed Tomography (CBCT) and Tomosynthesis. However, a major drawback of digital X-ray technology is cost (production, maintenance and replacement). Flat panel detectors utilize advanced semiconductor components that require costly cleanroom manufacturing processes, which leads to high eventual costs for the final product, but not only: it also leads to high maintenance/repairing costs. The complexity of the technology also mandates repair by specialized personnel, who are not commonly available in low and middle-income countries (LMICs). Additionally, we have received multiple reports from health personnel in Sub-Saharan countries, that the quality of images made with flat-panel detectors is degraded when temperature increases above a certain level. Some radiographers even reported that due to this effect, they stopped performing radiographies when temperatures were above "27 or so" degrees Celsius. Flat-panels seem to be ill adapted to harsh environments and, if damaged, the whole detector should be replaced by a specialist at considerable cost.

The alternative to flat-panels, are the multi-camera array detectors, such as the IONA from TeleOptic or the Naomi from RFSystems. These systems generate the images using a scintillator and an array of cameras. Unlike flat panels, the multi-camera detectors do not use large, brittle semiconductor substrates but small image sensors. Therefore, they are more robust and can be produced at a lower cost. However, their lower optical coupling lead to an increase in image noise and a decrease in detective efficiency [1-3].

While the containment and reduction of healthcare costs is an intense topic worldwide, in LMICs the problem is considerably more serious. According to the World Health Organization (WHO) figures, more than two thirds of the world's population does not have access to essential x-ray imaging equipment [4]. Too often in developing countries, patients die of trivial problems, which, due to a lack of access to diagnosis, take dramatic proportions. Road accidents, tuberculosis, and complications from childhood pneumonia are recurrent examples of pathologies causing complications that could have been prevented with functional and efficient x-ray imaging services [5].

Among the few medical X-ray systems available in developing countries, the overwhelming majority are still film-based, which leads to very high operating costs and often yields poor image quality. Modern digital X-ray systems are not widely available in these countries. This is due to the mismatch between existing solutions and the local context: the high cost of equipment that contributes to the increase in the costs of healthcare is one key issue. Others are related to the lack of quality infrastructure i.e. lack of stable electrical power, which tend to cause frequent and extended downtime. The scarcity of trained personnel to use, maintain and repair the devices when they need it, are yet another reason of a reduced lifespan [6]. Finally and very importantly the harsh environment, which involves high levels of humidity and high temperatures causes high failure rates. The performance and lifespan of digital X-ray equipment, detectors especially, are prone to be compromised at high humidity rate and temperatures [6].

The GlobalDiagnostiX project aims to develop, in partnership with local actors in Cameroon, a digital, ultra-robust, and affordable radiological x-ray equipment adapted to the needs and constraints of low and middle-income countries. This project, while lead in the frame of a large academic alliance, involves not only developing the technology itself but also a sustainable business model. The methodology [7] relies on three pillars: 1) cooperation and co-creation with local stakeholders in Cameroon, 2) insterdisciplinarity, with participation of engineers, radiologists, radiographers, anthropologists, designers etc. and 3) entrepreneurship as the output of the academic work provided the basis for an award-winning start-up company. In this frame, we have developed a novel, low-cost X-ray imaging system, based on a multi-camera architecture, capable of providing a diagnostic image quality in harsh environments. The developed system implements innovative real-time hardware electronics that comprises multiple Complementary Metal Oxide Semiconductor (CMOS) cameras, providing each a partial image that are unwarped, enhanced and combined through parallel processing units (FPGA) and a specialized software.

The proposed X-ray detector is based on indirect conversion [Fig 1]: the x-ray photons hit a scintillator screen that emits multiple visible photons upon absorption. These secondary photons form a visible image that is then captured by an array of cameras. The camera array uses Off-the-shelf CMOS image sensors and optical lenses to reduce fabrication costs and to ensure easy access to replacement parts. Additionally, the implemented CMOS sensors are less sensitive to temperature and can operate at higher temperatures than other image sensor technologies [8]. Finally, dedicated electronic and computing units such as field programmable gate array (FPGA) chips collect the data from the sensors and send a reconstructed image to an external computer for visualization.

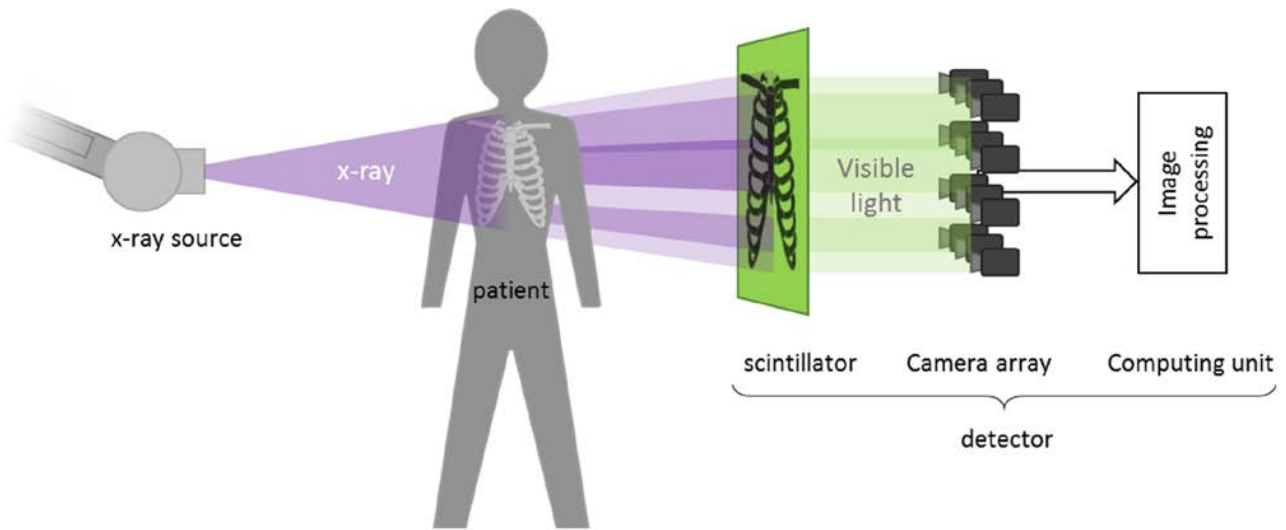

**Fig 1.** Schematic of the detector's indirect x-ray conversion chain.

During the development phase, different commercial and customized elements were tested, for section of the conversion chain, in order to define an optimum set of components to achieve state-of-the-art clinical images under harsh environment conditions. Each of the detector components, the development and characterization process and the implemented system architecture are described in detail in the following sections.

# MATERIALS & METHODS

## Characterized components

### Scintillators

The scintillator is a key element in the conversion chain. It converts X-rays to visible photons with high efficiency and very low lateral scattering. When an X-ray photon hits an atom of the scintillating material, an electron from the inner orbital is ejected. Surroundings electrons with higher energy will take the empty place by releasing the excess of energy with multiple photons at lower energy (in the range of 2-2.5 eV, visible spectra) [9,10].

Indirect conversion digital X-ray detectors for medical applications often implies one of two types of scintillator materials: Cesium Iodide (CsI:Ti) or Gadolinium Oxysulfide ($Gd_2O_2S$:Tb). Both have advantages and disadvantages regarding light throughput, resolution, price, resistance to temperature and humidity, etc. In order to identify the best X-ray conversion device, in terms of performance and cost, several scintillators (CsI:Ti and $Gd_2O_2S$:Tb) from different manufacturers were analyzed. After a preselection, based on the active area, thickness and resolution, seven models of commercial scintillators were acquired for characterization. Table 1 lists the selected scintillators with their main features.

**Table 1. Characterized scintillators**

| Scintillator | Origin | Technology | Size [mm] | Thickness |
|---|---|---|---|---|
| SC1(CI) | Japan | Csi:Ti | 430x430 | 400 um |
| SC2(CI) | China | Csi:Ti | 430x430 | 400 um |
| SC3(GOS) | Japan | Gd2O2S:Tb | 430x430 | 140 um |
| SC4(GOS) | Japan | Gd2O2S:Tb | 430x430 | 208 um |
| SC5(GOS) | China | Gd2O2S:Tb | 430x430 | 390 um |
| SC6(GOS) | UK | Gd2O2S:Tb | 430x430 | 250 um |
| SC7(GOS) | UK | Gd2O2S:Tb | 430x430 | 250 um |

*Characterized commercial scintillators with their main features.*

# Lenses

The lens has the role of collecting the visible photons and makes them converge to an image sensor, with as little optical aberrations as possible. Its importance in the conversion chain is not to be neglected: a low quality lens (i.e. aberration, low transmission rate, etc.) leads to bad optical coupling, low modulation transfer function (MTF) and detective quantum efficiency (DQE). The specifications of the lens (i.e. focal length, aperture) define the magnification factor (or field of view), which has a direct impact on the minimum distance between the scintillator and the sensor to have a good overlap between the sub-images.

Although commercial lenses have a multitude of different mount designs in the machine vision industry, there are three widely used formats, i.e. C-mount, CS-mount and S-mount (also called M12). After a careful analysis of the different lens mountings, the S-mount lenses were selected for the detector design due to their small size, flexibility, low maintenance and cost efficiency. Once the mount type was selected, nine of the most suitable lenses were purchased for characterization (Table 2).

**Table 2. Characterized lenses**

| Lens | Origin  | Focal length [mm] | F-stop | Sensor type | Working distance [mm] |
|------|---------|-------------------|--------|-------------|-----------------------|
| L1   | Germany | 6                 | 1.6    | 1/2.5       | 102                   |
| L2   | Germany | 6                 | 1.6    | 1/2         | 100                   |
| L3   | China   | 6                 | 1.2    | 1/2.7       | 97                    |
| L4   | China   | 2.8               | 1.2    | 1/2.7       | 55                    |
| L5   | China   | 3.6               | 1.2    | 1/2.7       | 68                    |
| L6   | Germany | 6                 | 1.6    | 1/2         | 100                   |
| L7   | Germany | 16                | 1.6    | 1/2         | 267                   |
| L8   | Germany | 6                 | 1.2    | 1/3         | 105                   |
| L9   | Germany | 8                 | 1.2    | 1/3         | 138                   |

*Characterized lenses with their main features.*

Since the lenses present different specifications (e.g., focal length) and design, the field of view is different for each lens. Therefore, the working distance for each lens was set to have a similar image size on the scintillator of that of the reference configuration, i.e. L2 lens @100mm. This allows to compare the lenses efficiency with a fixed camera density for a 43x43cm$^2$ detector which is an usual size for a general purpose medical detector. The determined working distances of all lenses are summarized in Table 2, last column.

## Image sensors

Image sensors are semiconductor chips that embed an array of photodiode and some integrated electronic to drive and control this array. Compared to the Thin-Film-Transistor (TFT) technology used in X-ray flat panel detectors, CMOS technology benefits of silicon substrates offering the best performances and state-of-the-art semiconductor technologies.

By implementing an indirect-conversion architecture based on CMOS imaging sensors, we can profit from the widely available commercial sensors and lenses, leading to a more cost-effective solution. Moreover, the modularity of the system reduces the maintenance, the repairing complexity and costs, as only a component of the device can be replaced in case of failure instead of the full detector. On the other hand, the active area of the CMOS sensor is typically much smaller than the active area from flat panel sensors. Therefore, lenses are used for projection and demagnification of the scintillator image. This reduces the optical coupling leading to a reduction of the final DQE.

The market was screened for cost efficient CMOS image sensors with suitable resolution. Color image sensors use Bayer filters to discriminate photons of certain wavelength (color). Since the scintillator light is monochromatic (green), only 50% of the sensor pixel will be able to capture the generated image, i.e. the other 50% of the pixels (with red and blue filters) will not be able to capture any light. This significantly reduces the light transmission and therefore the overall sensitivity of the detector. Therefore, the monochrome image sensors represent the best fit for this application. On the other hand, color sensors are more demanded and thus usually more readily available and for a better price. Therefore, a set of 2 monochrome and 2 color sensors were selected, based on the manufacturer specifications, for an extensive comparison according to the EMVA 1288 3.1 standard [11]. The key selection parameters were signal-to-noise ratio (SNR) and sensitivity performance in low light conditions. The selected image sensors, and their main features, are presented in Table 3.

**Table 3**. **Selected sensors.**

| Image Sensor | Origin | Technology | Color | Resolution | Size |
|---|---|---|---|---|---|
| IS1 | Japan | Back Illuminated | Monochrome | 1936x1096 | 1/2.8" |
| IS2 | Japan | Back Illuminated | Color | 1936x1096 | 1/2.8" |
| IS3 | USA | Front Illuminated | Monochrome | 1280x960 | 1/3" |
| IS4 | USA | Front Illuminated | Color | 1928x1088 | 1/2.7" |

*Characterized sensors with their main features.*

The characterization of the IS3 Monochrome and IS4 Color sensors was performed using the evaluation boards provided by the manufacturer. For the IS1 (Monochrome) and IS2 (Color) sensors, commercial cameras were used.

## Sensor shield

Scintillators have an absorption rate that varies between 32% and 75% depending on the type of scintillator and the X-ray spectra. The residual (not absorbed) X-ray photons interact with all the components behind the scintillator, i.e. the mechanical support, the lenses, the image sensor and the neighboring electronics.

X-ray tests show that the impact of the residual X-ray photons on the image sensor is non-negligible. Part of the residual X-ray photons are absorbed by the image sensor pixels, resulting in very high pixel values in some localized areas (denominated "direct hits") as exemplified in **Fig 2**. From a medical point of view, the number of direct hits can highly affect the ability to perform a good medical diagnostic, i.e. the white spot artefacts can hide small features, thus valuable information.

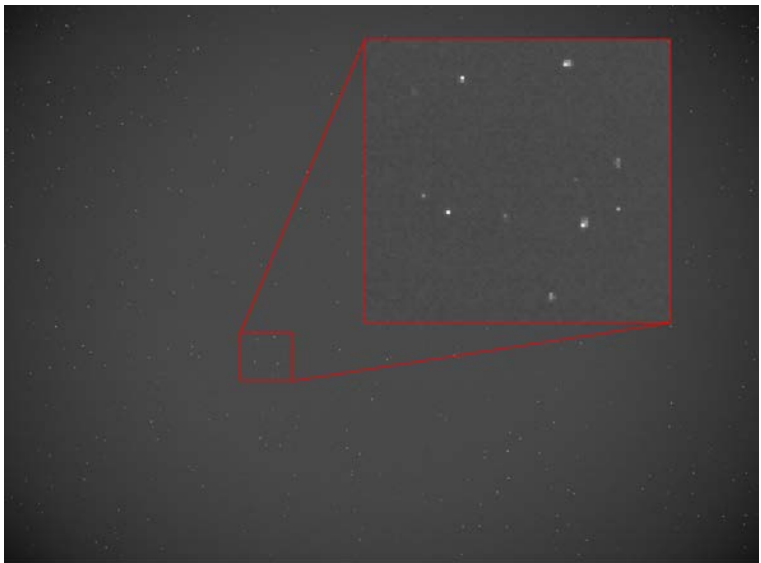

**Fig 2.** Examples of direct hits (white spots) on a test image taken with no lead glass (@ 80kV 30mAs).

A software solution can be applied to filter out the white spots by detecting and replacing the affected pixels with neighboring pixel values. However, correcting a very high number of direct hits may alter the image and lose too much information for a correct medical diagnosis. Therefore, to avoid losing information, a hardware solution must be added to protect the sensor, in addition to a software-based image correction.

Lead glass has a good X-ray photons absorbance and it is transparent to visible light. Therefore, placing a lead glass shield between the scintillator and the camera can absorb the residual X-rays while letting the visible light pass through. X-ray absorbance is linked to the concentration of Pb in the glass, usually given in "Pb equivalent thickness". However, these Pb atoms tends to tint the glass yellow, thus filtering some of the visible light. In order to reach a good trade-off between X-ray absorbance and light transmittance, different tests were performed with several Pb equivalent thickness glasses.

# Image quality assessment

## X-ray characterization

The assessment of image quality required the measurement of the MTF, the noise power spectrum (NPS) and the DQE [12].

**X-ray imaging setup and beam qualities:** The detector was characterized for the standard beam quality RQA 5 (70 kV – additional filtration of 21 mm Al at the tube exit) defined in the IEC 62220-1 document [13]. A scintillator was positioned 100 mm in front of the sensors and a lead glass of equivalent thickness 1.5 mm lead was interposed behind the scintillator to avoid direct detections by the pixels. The distance source-to-scintillator was set at 132 cm. Pre-processed digital images in a raw 16-bit format were used. The exposure time was varied to give different detector air kerma (DAK). All the images were obtained without anti-scatter grid. Air kerma measurements were made with a Radcal 9015 dosimeter with a 6 cm$^3$ ionizing chamber. A standard air kerma value of 2.5 µGy was taken as the reference dose level at the detector. Several other target DAK between 0.59 and 18.7 µGy were analyzed.

**System response:** Uniform images acquired at different DAK were used to measure the relationship between the mean pixel value (PV) and the DAK. Regions of interest (ROI) were selected at the center of each image for calculating the mean pixel value. The system response curve was fitted using a linear function:

$$PV = a + b \cdot DAK ,\tag{1}$$

Where a and b are fitted coefficients. The response curve was used to express the image pixel values into DAK levels for the MTF and NPS calculations.

**Modulation Transfer Function (MTF):** The MTF assesses the spatial resolution of the imaging system. The limit in resolution is often given as the spatial frequency at which the MTF value is equal to 5% [14]. According the Swiss Regulation RO 1998 1084 Annex 12.8, a minimum of 2.8 lp/mm is required for X-ray detectors intended to medical applications.

For this study, a tungsten sharp edge was imaged to produce the edge spread function (ESF). The derivative of the ESF gave the line spread function (LSF), the impulse response of the imaging system. The MTF is the magnitude of the Fourier transform of the LSF [15].

Unless otherwise stated, all images for the MTF measurements were acquired using a RQA5 beam with a SC5(GOS) scintillator, L2 lenses and the monochromatic sensor IS1, at a source-to-scintillator distance of 140mm. A lead glass shield of 1.5mm Pb equivalent was inserted between the scintillator and the camera to prevent the residual X-rays from reaching the camera.

**Noise Power Spectrum (NPS):** The NPS describes the frequency content of the image noise. 2D NPS are the magnitude squared of the Fourier transform of a homogenous region of interest (ROI) on the image that contains only noise:

$$NPS(f_x, f_y) = \frac{1}{A} \langle |\iint_A (d(x,y) - \bar{d}) e^{-i2\pi(xf_x + yf_y)} dxdy|^2 \rangle, \qquad (2)$$

Where $A$ is the area of the image, $d(x,y)$ the pixel value at position (x,y), $f_x$ and $f_y$ are spatial frequencies in the x- and y-directions, respectively, and $\bar{d}$ is the mean pixel value in the ROI. The NPS was calculated from three identical homogenous images for each detector air kerma (DAK). The normalized noise power spectrum $NNSP(f_x, f_y)$ is the NPS normalized by the square of the mean pixel value:

$$NNPS(f_x, f_y) = \frac{NPS(f_x, f_y)}{\bar{d}^2}, \qquad (3)$$

1D NPS curves are radial averages of 2D NPS, excluding the 0° and 90° axial values.

**Detective Quantum Efficiency (DQE):** The DQE quantifies the efficiency of the detector to convert incident X-ray photons into digital information. High DQE value allows using less patient dose for the same image quality. Consequently, optimizing the DQE is a major concern in the design of X-ray detectors. The DQE is the ratio between the output signal-to-noise ratio squared ($SNR_{out}^2$) and the input signal-to-noise ratio squared ($SNR_{in}^2$) in the spatial frequency space [13-16]:

$$DQE = \frac{SNR_{out}^2}{SNR_{in}^2}, \qquad (4)$$

The DQE is comprised between 0 and 1, 1 being a lossless detector. The DQE is proportional to the square of the MTF, and inversely proportional to the NNPS and X-ray photon fluence (Q).

$$DQE(f_x, f_y) = \frac{MTF(f_x,f_y)^2}{NNPS(f_x,f_y) \cdot Q}, \tag{5}$$

The photon fluence (Q) is given by the product of the DAK and the X-ray fluence per unit DAK ($\varphi$):

$$Q = DAK \cdot \varphi \left[\frac{\#photons}{mm^2}\right], \tag{6}$$

The spectral X-ray fluence per DAK is the number of X-ray photons per surface unit per dose unit [$\#photons/(mm^2 \cdot \mu Gy)$], and depends on the X-ray beam spectrum (kV and filtration) [13].

**Spectroscopy:** In order to precisely identify the requirements of the imaging sensor, spectroscopy measurements were performed in all the scintillators to determine their emitted light spectrum. Due to the low sensitivity of the spectrometer, high voltage (140kV) and high dose (160mAs) were applied, and no filter was added at the output of the tube during these tests.

## Visible light characterization

To study the image sensors performance in detail without interferences from external systems (e.g. scintillator, x-ray source, etc.), the image sensors and the developed detection system were tested under visible light: All image sensors were characterized without optics in darkroom conditions. A 525 nm green LED - a wavelength as close as possible to that of the scintillator - was selected as light source. The light source was diffused with an integrating sphere, to achieve the most homogeneous illumination of the sensor. The distance between the sphere output and the sensor plane was set to 100 mm. The irradiance of the source was measured with a calibrated photodiode and set to 1.0 µW/cm2 (at the sensor plane). Each sensor was set at the smallest gain that achieved the highest pixel reading at saturation, to ensure the full dynamic range is used. A schematic of the implemented measurement setup is shown in **Fig 3**.

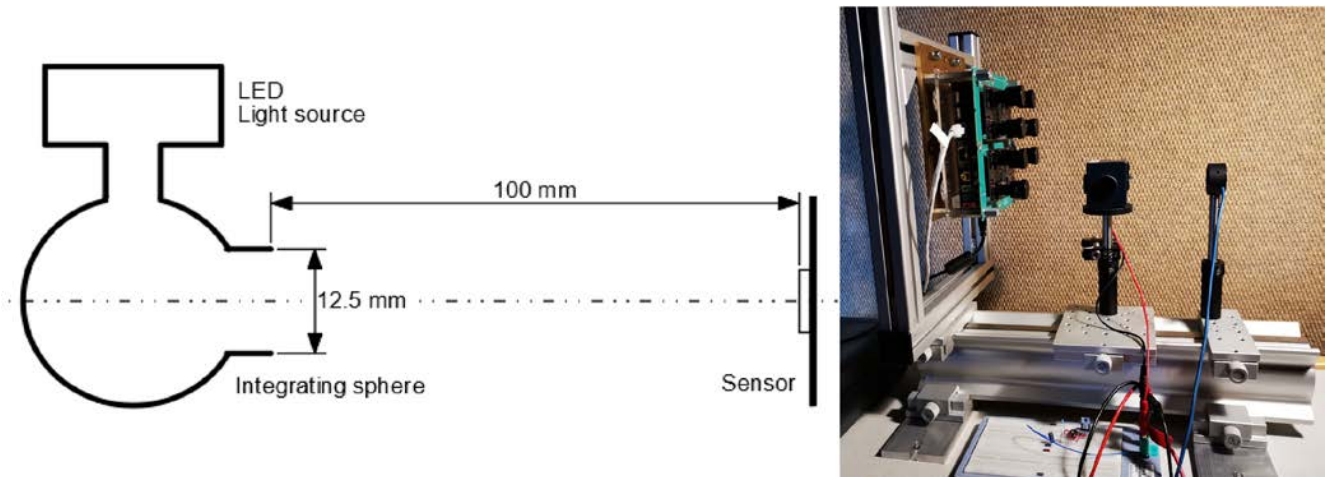

**Fig 3.** Schematic and picture of the image sensors' characterization setup.

Based on the EMVA1288 standard, different series of images were shot with varying exposition time and constant illumination. Sets of two images were captured with exposure increasing linearly up to pixel saturation with 50 steps, a first series was shot with the light source ON and the second in total darkness. Subsequently, two additional series of 50 images were shot with illumination and exposure times corresponding to 50% of the pixels saturation value. Again, the first series was shot with the light source ON and the second in the dark. Finally, the data from the captured images was processed in order to extract the sensor's intrinsic parameters, e.g. quantum efficiency (QE), gain (K), signal to noise ratio ($SNR_{MAX}$) and dynamic range.

# Components' characterization results

## Scintillators

The selected scintillators were characterized using the methods described in section "Image quality assessment". It is to be noted that the DQE and MTF were measured with non-optimized optical elements (lens) and the results are meant for relative comparison only.

## *MTF*

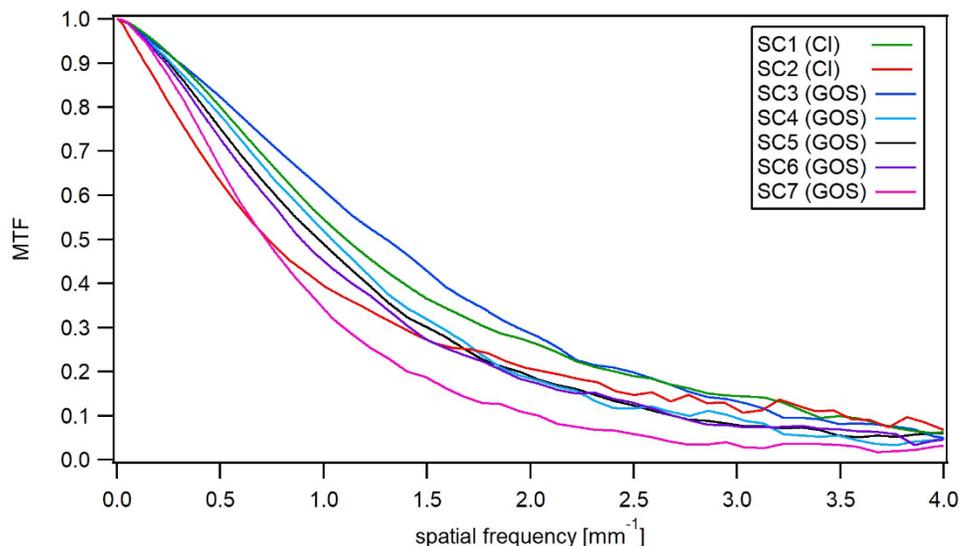

**Fig 4.** MTF curves of all tested scintillators under RQA5, 1.5mm Pb equivalent lead glass.

As a result of its intrinsic micropillar structure, CsI:Ti scintillators are known to have very good spatial resolution in comparison to other types of scintillator (e.g. $Gd_2O_2S$:Tb). This is reflected in the excellent behavior observed with the SC1(CI) scintillator, leading to high MTF values at low and high spatial frequencies with the best measured cut-off frequency (4.52 lp/mm). The other CsI:Ti scintillator, the SC2(CI), shows relatively low MTF values at low spatial frequencies but recovers after 2 lp/mm, reaching a cut-off frequency of 4.07 lp/mm.

The $Gd_2O_2S$:Tb scintillators were expected to show lower MTF levels than the CsI:Ti scintillators due to their powder structure. Surprisingly, the SC3(GOS) obtained the highest MTF at low spatial frequencies. However, its MTF rapidly decreases above 2 lp/mm, which leads to a cut-off frequency of 3.99 lp/mm. The SC5(GOS) showed an average behavior at low and medium spatial frequencies. Nevertheless, its MTF remains almost flat at high frequency leading to the second highest high cut-off frequency from the measured scintillators, i.e. 4.28lp/mm. The SC7(GOS) has very poor MTF levels, probably due to its higher thickness compared to the other $Gd_2O_2S$:Tb scintillators. It only reached a 2.6 lp/mm cut-off frequency, below the minimum 2.8 lp/mm required for film-screen X-ray detectors intended to radiological applications.

## DQE

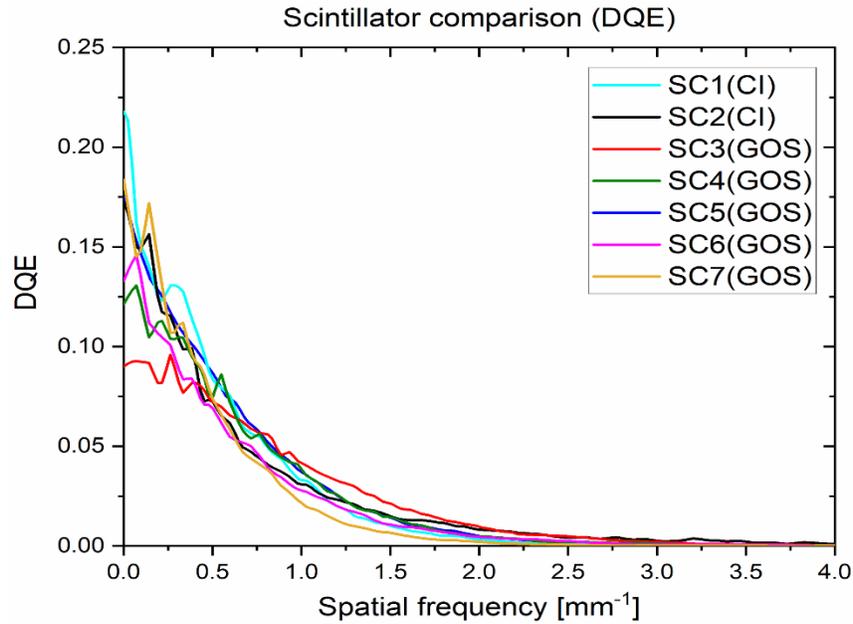

**Fig 5.** DQE curve comparison of the tested scintillators @18uGy.

Due to its high thickness and absorption rate, the SC1(CI) Scintillator has the highest low-frequency DQE. The SC2(CI) shows relatively poor performance for a CsI:Ti scintillator. Surprisingly, the SC5(GOS), a low cost Gd2O2S:Tb scintillator, produced the third highest DQE levels amongst the measured scintillators. The SC7(GOS) outperforms the SC2(CI) and the SC5(GOS) with a DQE(0) of 18.4%. The high DQE(0) levels of the SC5(GOS) and the SC7(GOS) can be explained by their thicker active layer, i.e. higher absorption. However, higher thickness can lead to poorer MTF, due to more lateral diffusion in the scintillating layer. This is especially critical in the case of the SC6(GOS) and SC7(GOS).

## NNPS vs. temperature

The X-ray detector will operate in harsh environments, where ambient temperature can reach up to 45°C with 100% relative humidity. The Gd2O2S:Tb scintillators are known to be very robust and stable, however, the CsI:Ti can suffer of image quality loss at high temperatures and humidity. The impact of the temperature on image noise (NNPS) was measured for two scintillators at different temperatures, i.e. the SC3(GOS) (Gd2O2S:Tb) and the SC2(CI) (CsI:Ti). The resulting NNPS curves are shown in Fig 6 .

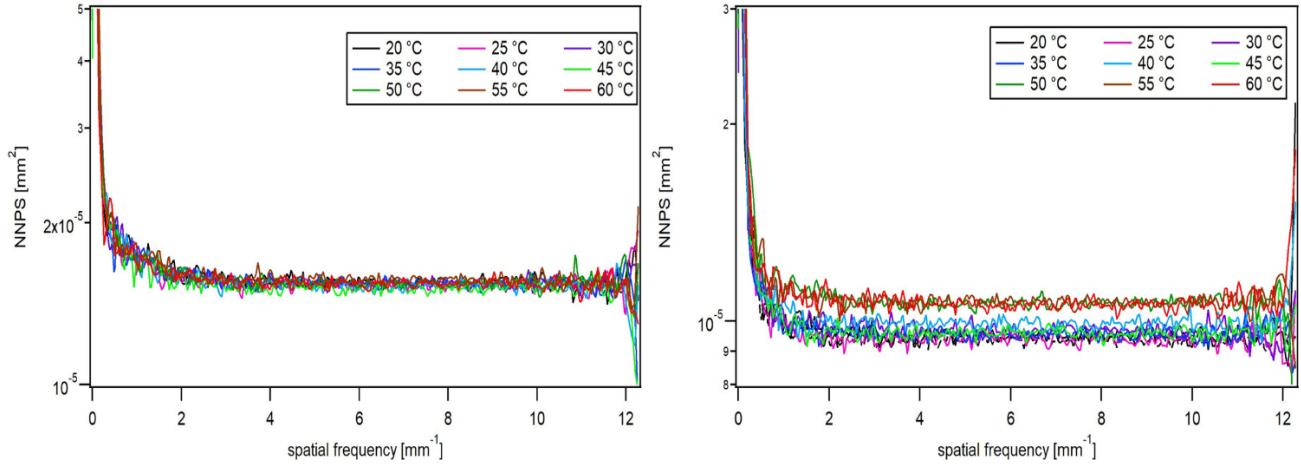

**Fig 6.** NNPS curves for the SC3(GOS) (left) and the SC2(CI) (right) at different temperatures.

As expected, the SC3(GOS) exhibits no change over the tested temperature range. Conversely, the SC2(CI) exhibits an increase of the NNPS by about 10% at high temperature (>45ºC). Since the NNPS is inversely proportional to the DQE, this would lead to a 10% decrease in DQE when used at a temperature over 45°C.

## Spectroscopy

The CsI:Ti scintillators have a very broad emission bandwidth (from approximately 470nm to 640nm), thus the peak value is lower compared to $Gd_2O_2S$:Tb. All $Gd_2O_2S$:Tb scintillators exhibit a high peak emission at around 545nm and few other peaks at around 490nm, 580nm and 620nm ( Fig 7 ).

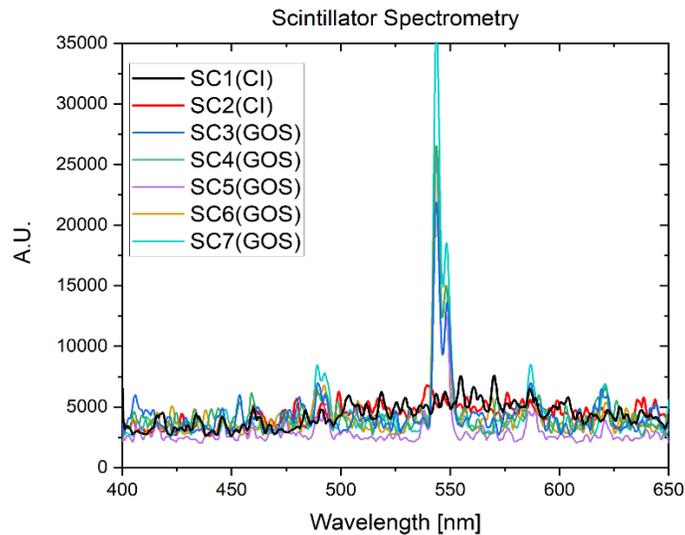

**Fig 7.** Emitted light spectra of the tested scintillators.

## Performance and cost analysis

The SC1(CI) showed the best image quality amongst the tested scintillators. However, it has strong drawbacks inherent to CsI:Ti scintillators, e.g. high price, performance decrease at high temperature and degradation due to humidity [17] (CsI is slightly hygroscopic, i.e. tends to absorb moisture from the air [18]). This presents a big challenge for their implementation in an X-ray detector intended for low-income tropical countries. The other CsI:Ti scintillator, the SC2(CI), exhibited the same drawbacks as the SC1(CI) with a lower image quality.

SC6(GOS) and SC7(GOS) scintillators are cost-effective but showed very poor performances with the exception of a very good DQE(0) for the SC7(GOS). However, this is at the expense of the MTF, which leads to a cutoff frequency below the minimum required for film-screen X-ray detectors intended to medical applications. The SC3(GOS) and SC4(GOS) scintillators (both from the same manufacturer) were the most expensive amongst the tested $Gd_2O_2S$:Tb scintillators, however their performance, both in MTF and DQE, was below average.

The SC5(GOS) outperformed all $Gd_2O_2S$:Tb scintillators in MTF and DQE (except for the SC7(GOS) at low frequency (DQE(0)). Also, it does not suffer from image quality loss due to high temperature and humidity ($Gd_2O_2S$:Tb is not hygroscopic [19]). Moreover, it is the most cost efficient of all tested scintillators. This makes it a promising candidate for an X-ray detector intended for low-income tropical countries.

## Lenses

### MTF

The measured MTF for the different lenses are presented in Fig 8. The MTF were measured at the center of the image, where the focus of the lens is optimal. Due to Seidel's curvature of field effect, edges of an image tends to be slightly out-of-focus when the center is in focus. This effect depends on the lens characteristics, size of the object and image, and the focal distance. To determine the effect of the out-of-focus aberration, the MTF at the edge of the images was calculated and compared to the center values. As expected, the MTF measured in the edge is lower than the MTF in the center, however, the maximal variations are low (<0.03).

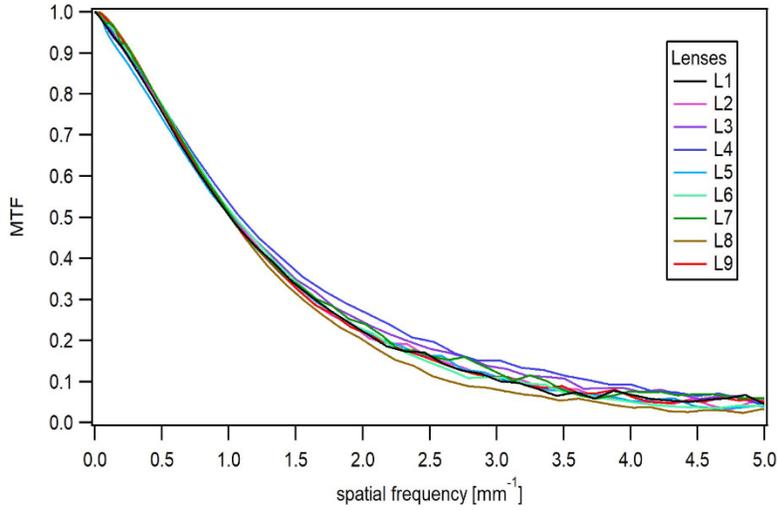

**Fig 8.** MTF comparison of all lenses, with the SC5(GOS) scintillator and IS1 sensor, RQA5 beam (70kV and 20mAs).

All the characterized lenses show a similar MTF behavior. The 1270XX12MP-M12 lenses slightly outperforms the others in terms of spatial resolution, which is surprising given their low price. On the other hand, the German L8 lens has the lowest spatial resolution. Nevertheless, its cut-off frequency (3.75 lp/mm) is still much higher than the minimum required.

## DQE

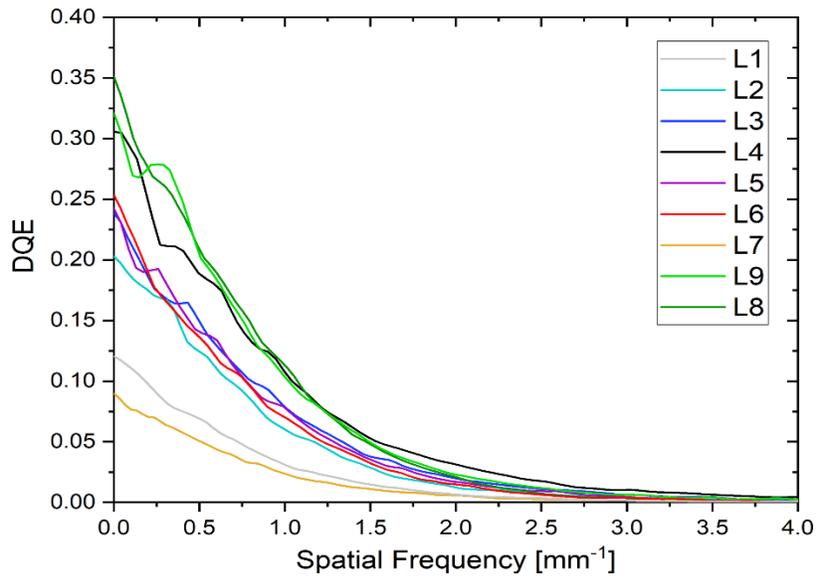

**Fig.9.** DQE comparison of all lenses, with the SC5(GOS) scintillator and the IS1 sensor, RQA5 beam (70kV and 3.10mAs, 2.5uGy)

*DQE(0) vs. focal length*: The higher the focal length, the lower the DQE. This is explained by the fact that with lower focal length, the camera is closer to the scintillator (for imaging the same area), thus more photons reach the image sensor. Nevertheless, low focal length usually comes with more geometrical aberration for such small size, low cost lenses, i.e. little room for aberration correction. Therefore, a trade-off between light collection and image distortion must be carefully defined.

*DQE(0) vs. F-stop:* The F-stop is linked to the numerical aperture and the lens entrance pupil. The wider the entrance pupil (or the lower F-stop), the more light can pass through the lens. Therefore, the lower the F-stop, the higher the DQE. However, a higher quality lens with higher F-stop can have a higher DQE(0) than a lower quality lens with lower F-stop. Such is the case of the L6 (f 1.6) which has a higher DQE(0) than two of the lenses (L3 and L5) with lower F-stop (1.2). This shows that, despite the theoretical features described in the datasheet, the lens performances are highly dependent on the quality of manufacturing.

## Performance and cost analysis

After analyzing the experimental results, lenses with focal length 6 mm represent the best tradeoff between geometric distortion, field of view, and light throughput. In addition, 6 mm focal length lenses are very common in the market. Therefore, they are easy to obtain, there are many models to choose from and their prices are low in comparison to more specialized optics.

The L8 lens is the best candidate for the detector system. It gave the best zero-frequency DQE levels from the measured lenses at an acceptable cost. Its main drawback is its low MTF, which however remains higher than the minimum required.

## Image sensors

The measured sensitivity and SNR plots for all characterized sensors are presented in **Fig 10**. For the color sensors, only the green channel was considered (50% of the pixels). Therefore, to account for the loss of sensitivity due to the Bayer filter, the pixel surface was considered be that of two pixels. Table 4 summarizes all the characterized sensors and measured metrics.

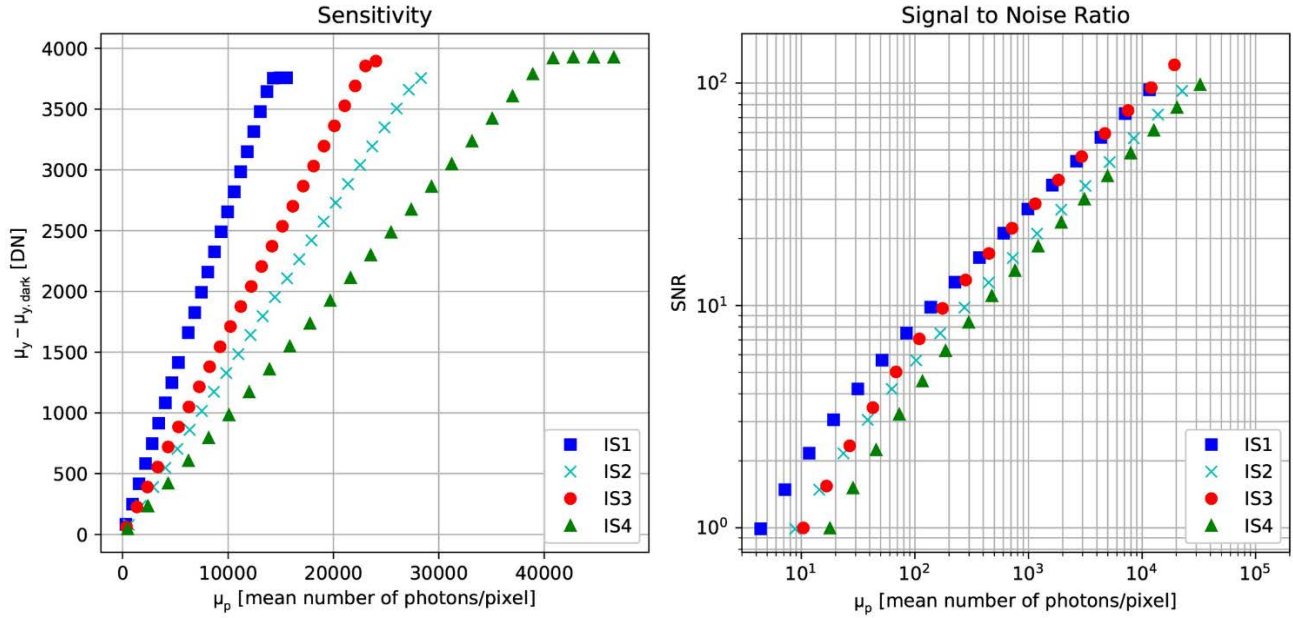

**Fig 10.** Image sensors characterization results for sensitivity and signal to noise ratio (SNR). $\mu_y$ is mean value of light image in digital value (DN) and $\mu_{y\,dark}$ the mean value of dark image in DN to remove in order to make black equal to 0 (removal of pedestal level).

**Table 4**. **Characterized sensors' results.**

| Parameter | Unit | IS1 | IS2 | IS3 | IS4 |
|---|---|---|---|---|---|
| Pixel size | µm | 2.90 | 2 x 2.90 | 3.75 | 2 x 3.00 |
| Quantum efficiency | % | 75.1 | 37.8 | 76.0 | 29.9 |
| System gain | DN/e⁻ | 0.36 | 0.36 | 0.22 | 0.33 |
| Temporal dark noise | e⁻ | 2.67 | 2.71 | 7.30 | 4.74 |
| Absolute sensitivity threshold | p | 4.38 | 8.81 | 10.4 | 17.8 |

*Main data results from the sensors' characterization*

## Sensitivity

The sensor's sensitivity is given by its mean pixel response to light (in digital number (DN)). The fixed irradiation, wavelength and exposure time were used to compute the number of photons per pixels for each measured point. All sensors showed linear responses up to saturation. The IS1 (**Fig 10** (Left)- dark blue squares) shows the best sensitivity behavior from the measured sensors. This can be explained by the "back-side illuminated" technology used in its design. In "back-side illuminated" sensors, the photodiode is located over the integrated circuit and metal lines (opposite to traditional C-MOS sensors), this reduces the light scattering and reflection, thus more photons reach the photodiode (improving the sensitivity and reducing the noise). The IS2 sensor uses the same technology as the IS1, however, its results are below the monochrome version. This is due to the fact that only 50% of its pixels are able to detect green light (Bayer filter), therefore, half of the photons are lost.

### SNR

The signal-to-noise ratio (SNR) is computed by dividing the mean value of an image with homogeneous illumination by its standard deviation. EMVA 1288 standard proposes a camera model with a single internal noise source that is the sum of dark and quantization noise. The number of photons also fluctuate statistically (shot noise), this phenomenon is not negligible for a small number of photons.

Results in low light conditions, with a low number of photons, are of most interest for the X-ray detection application. Since it would be difficult to properly measure the SNR with an illumination of only some photons, the EMVA 1288 standard proposes a computation to extrapolate the SNR based on the fitted camera model. The plotted curves correspond to this extrapolation and make the absolute sensitivity threshold visible i.e. absolute sensitivity threshold corresponds to the point where the noise is equal to the signal and thus the smallest amount of photons detectable by the camera. Again, the IS1 Monochrome (**Fig 10** (Right) – dark blue squares) shows the best behavior from the characterized sensors, especially in low light condition, thanks to its low-noise technology.

### Performance analysis

Multiple CMOS sensors were tested according to the EMVA1288 standard. Low light performance is the key factor in selecting the correct sensor for this application. Results showed that the new IS1 outperforms all other tested sensors in that regard. Therefore, it is a good candidate for the multi-camera array X-ray detector.

## Sensor shielding (Lead glass)

In order to reach a good trade-off between X-ray absorbance and light transmittance, different tests were performed with several Pb equivalent thickness glasses (from 0.5 to 3 mm). Since the lead glass absorption rate depends on the energy of the X-ray photons, three series of tests were performed at 70, 80 and 120kV. The detection and counting of the direct hits were done with a customized threshold algorithm.

Placing the lead glass shield in front of the sensor resulted in a dramatic reduction of the number of direct hits: The higher the Pb equivalent thickness the lower the number of direct hits, i.e. higher X-ray absorption rate, until 1.5 mm where it stabilizes in an average hits reduction of around 93%. The remaining hits can be attributed to the diffusing and back-scattering X-ray photons due to the setup.

The drawback of using a lead glass shield is a reduction in the light transmittance. Transmittance measurements, for a wavelength range from 500nm to 600nm, showed a light absorption of around 7% by the 0.5mm Pb equivalent thickness, which increases to 14 to 16% for the ≥1.5mm Pb equivalent thicknesses. This absorption will have a direct impact on the DQE, since the cameras will collect less light. However, this loss is considered acceptable in comparison to the benefit of having X-ray shielding to prevent residual X-ray photons to hit the sensor.

These results showed that, with a hit reduction of more than 93% and a reduction in light transmittance of less than 14%, the 1.5 mm Pb equivalent thickness glass is the best option to protect the sensors and other system electronic components from residual X-ray hits.

# IMAGE DETECTOR: DESIGN & ARCHITECTURE

The detector architecture consist of multiple multi-camera modules (4 camera per module), that comprise the CMOS image sensors, a computational unit (an FPGA) and memory (SRAM) to buffer the data. A master unit will collect the pre-processed images coming from these multi-camera modules and finalize the stitching. Based on the characterization results presented in the previous sections, the components showing the best compromise in terms of price, environment resistance and performances were selected (optimum configuration): the German lens L8, the monochromatic IS1 sensor and a lead glass of 1.5mm Pb equivalent thickness. With this configuration, and based on the tests results, an optimal scintillator to image sensor distance of 98.72 mm was determined. Taking into account the geometric distortion correction and the minimum overlap, this results in a matrix of 6 by 12 cameras (72 in total), as depicted in Fig 11 .

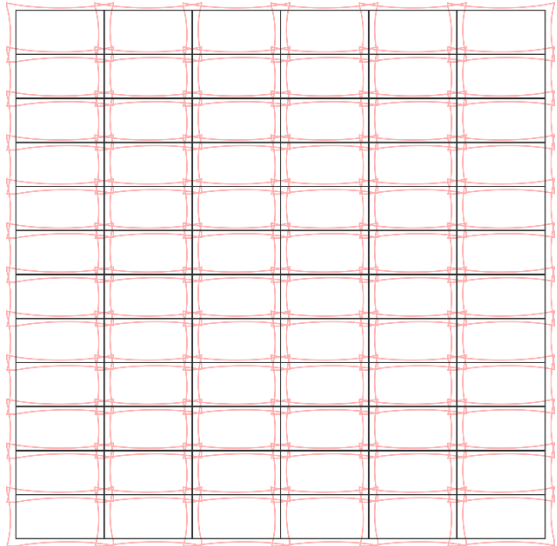

Fig 11.  Sub-image distribution for a 6 by 12 cameras .

## Multi-camera modules

An electronic imaging module embedding the selected image sensor was developed ( Fig 12 ). The 4xIS1 module embed four IS1 CMOS image sensors and their respective lens holders. Each 4-camera module uses 1 Intel FPGA Cyclone IV EP4CE22F17 chip to buffer the image in a small 32MB dedicated SRAM memory. Special care was taken to design low noise power supplies and lengths and impedance matched data paths to enhance signal integrity and minimize contribution of the board electronic noise to the sensors. The FPGA  includes some pre-processing algorithm to correct the flat image (for structured noise correction) and dark image (for electronic noise correction). The implemented topology enables to configure gains and exposures of many modules trough a single master unit, as well as triggering a simultaneous image capture from all sensors. The modules are designed to be mounted on a backplane circuit trough a high-speed connector system. This structure enables a region of the detector to be easily replaced for maintenance, if need be. Finally, the images are gathered by the FPGA master to be held at disposition of the control computer. Distributing the computing power to multiple modules makes the electronic architecture more complex. However, it also reduces the workload of the master unit and simplifies its interface.

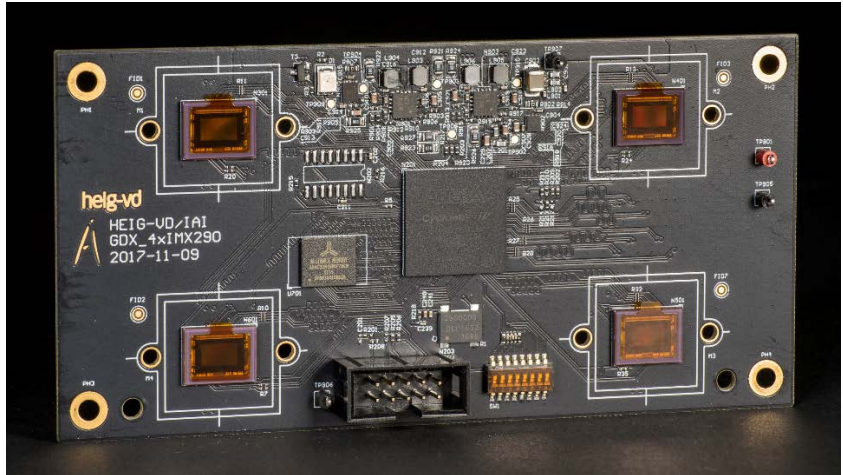

**Fig 12.** Module 4xIS1 (6 layers PCB with controlled impedance, BGA components)

## Master unit

A demoboard Terasic DE1-SoC was selected as master unit. The main task of the master unit is to interface all multi-camera modules by configuring the image sensors through SPI bus. The master unit also triggers all the cameras at the very same time in order to avoid any desynchronization between the sub-images. Once the pre-processed data from the multi-camera unit is retrieved into the master unit, it will finalize the image reconstruction in order to output a reconstructed image to an external PC via Ethernet protocol.

## Image processing

Each image sensor will produce a *raw image* corresponding to an imaged area of the scintillator, which is affected by a variety of intensity and geometric artifacts. Stitching is the operation which consists in assembling raw images into a final pre-processed image, which should not present any of these artifacts nor any observable assembly cues. Strictly speaking, basic (e.g., histogram equalization or edge enhancement) and advanced [20] image post-processing operations are not part of the stitching process and so they are beyond the scope of this paper. The overall stitching process is depicted in **Fig 13**.

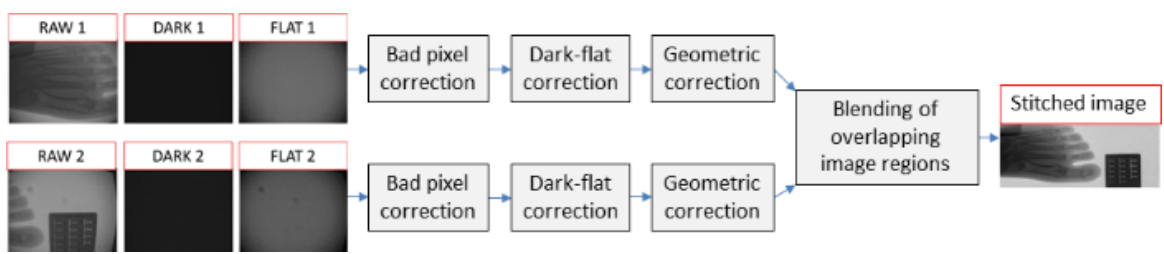

**Fig 13.** Overview of the stitching pipeline exemplified with a two cameras setup.

## *Bad pixels detection and removal*

Bad pixels refer to pixels with abnormal intensity such as dead (always off) or hot (always on) pixels. Such stuck pixels appear progressively over time due to sensor degradation and their position can be detected and permanently recorded. Hot pixels are also the result of X-ray direct hits not corrected by the sensor shielding solutions. By detecting the bad pixels, we can replace their intensity by a weighted average of their neighboring pixel intensities. Our bad pixel detection is based on the analysis of excessive deviations of pixels intensity in the raw and flat images compared to smoothed versions obtained by median filtering.

## **Dark-flat intensity correction**

Based on dark and flat images, raw images are corrected to get rid of the intensity inhomogeneity caused by the vignetting effect – mainly characterized by a light falloff far from the image center. Dark images are acquired in total darkness without X-ray emission while flat images are radiographs acquired without anything in front of the scintillator. The dark-flat correction is simple: $I = k(I_r - I_d)/(I_f - I_d)$, where $I_r$, $I_f$ and $I_d$ are the raw, flat and dark images. Parameter $k$ is a normalizing factor usually computed as the average of all pixel intensities of $(I_f - I_d)$. When correcting images from multiple sensors, the constant $k$ must be identical for all sensors to avoid global intensity inhomogeneity between corrected images $I$.

## **Geometric correction**

Each camera sensor $S_i$ is modeled as a pinhole camera characterized by the estimated intrinsic and extrinsic parameters [21]. Intrinsic parameters include the intrinsic matrix $K_i$ related to the lens characteristics (e.g. focal distance, pixel size) as well as tangential and radial distortion parameters $r_i$ modeling the non-linear deformation of the lens. Extrinsic parameters express the rigid transform $T_i = (R_i, t_i)$ from a world coordinate system $CS_w$ to a camera coordinate system $CS_i$, where $R_i$ and $t_i$ are the rotation and translation of the rigid transform.

Intrinsic calibration can be performed for each camera independently and with visible light using a calibration pattern. We selected a point cloud pattern (**Fig 14** a) from OpenCV library [22] that allows partial visibility of the pattern – a very useful feature given the small field of view and short focal distance of the camera sensors. In case of extrinsic calibration, we need to ensure that all cameras share the *same* world coordinate system $CS_w$ in order to estimate the spatial positioning and orientation of each camera with respect to each other. This is achieved by using a calibration pattern simultaneously imaged by all cameras and whose parts can be unambiguously detected. The acquisition must be done with *x-ray emission* as the conversion from x-rays to visible light takes place on the plane of the scintillator. Hence, the calibration pattern must not only offer partial detection but should be also X-ray "compatible" as well as easy to manufacture. In this context, we chose the random dots pattern from Oyamada et al. [23] composed of dots (**Fig 14** b) that can be easily drilled on an aluminum plate (**Fig 14** c), yielding highly contrasted features in the radiographs.

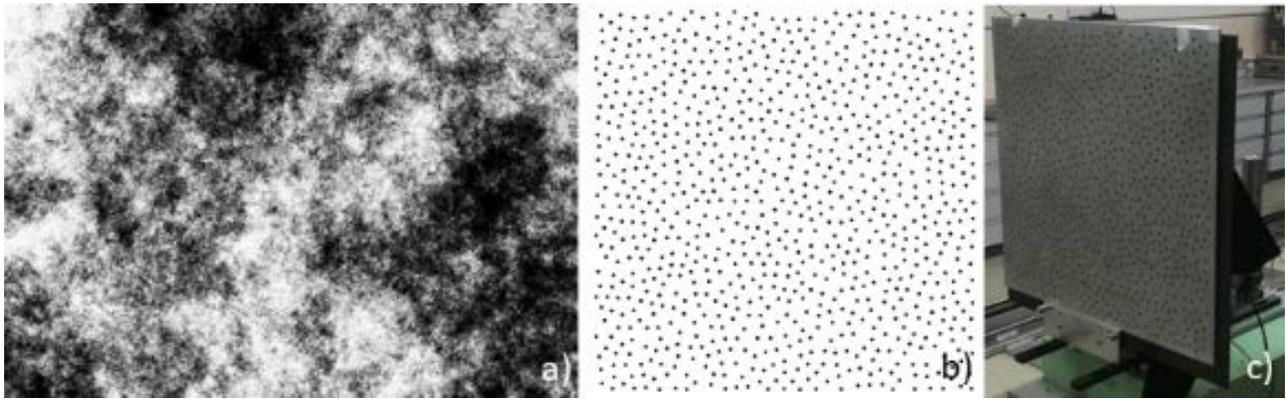

**Fig 14.** Cameral calibration patterns: **a)** point cloud pattern used for intrinsic calibration and **b)** random dots pattern [23] used to manufacture **c)** the extrinsic calibration aluminum plate.

Based on the estimated calibration parameters, a geometric correction process is applied on each sensor image $I_i$: (1) image undistortion produces rectified images without lens distortion using intrinsic parameters $K_i$ and $r_i$; (2) homography transformation converts the rectified image in the common world CS and resamples it to match a desired image resolution. The homography is computed from $K_i$ and extrinsic parameters $R_i$ and $t_i$. The geometric correction can be seen as an *unwarping* process using deformation maps, which can be preprocessed and applied very efficiently in hardware (e.g. FPGA).

## Blending

Unwarped images will present some overlapping regions as depicted in Fig 15 a. A successful dark-flat correction coupled with a gain compensation approach [24] will produce consistent image intensities across overlapping regions. This consistency avoids the need to identify frontiers in overlapping regions used to avoid so-called *seam* blending artifacts [25]. Similarly, advanced blending (e.g., multiband blending [26]) is not necessary and we can apply a simple weighted linear blending (also known as feathering) where weights are computed from distance maps of the unwarped image regions. As a result, the blending is simple, deterministic and very efficiently executed – yielding excellent stitching results as depicted in Fig 15 c.

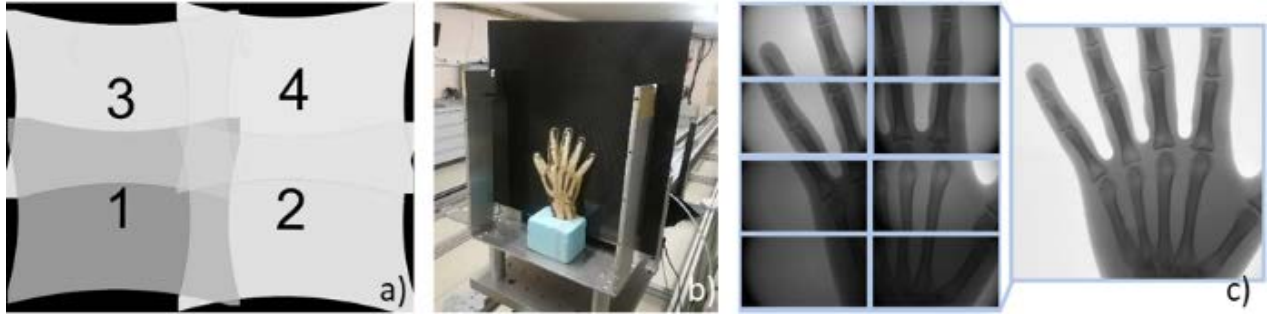

**Fig 15.** Blending and stitching: **a)** after geometric correction images will present overlapping areas that will be seamlessly blended into a final stitched image. An example of stitched image of a **b)** hand phantom and produced with an 8-sensors prototype is shown in **c)**.

# RESULTS

## Multi-camera module performance

Two types of multi-camera module were assembled (GDX_IS1_LR and GDX_IS1_SR) implementing linear and switching voltage regulators, respectively. Switching Voltage regulators have a higher power efficiency compared to linear regulators but they can add undesired electronic noise to the system. Both modules were tested with visible light and with X-rays.

### Visible light characterization

The visible light characterization of the developed multi-camera module was performed with the same method as the image sensor characterization ("Visible light characterization" Section), using the setup described in **Fig 3**. The first module to be characterized was the GDX_IS1_SR. In the initial measurements, a line pattern was visible on the image in low light conditions. This pattern was always present but randomly positioned, therefore jeopardized the DQE results. After thorough analysis, the switching voltage regulators were identified as the source of the high noise levels. The vertical spectrogram of Dark Signal Non-Uniformity (DSNU) showed a clear peak corresponding to the pattern spatial frequency in visible light (Fig 16 (right), red curve).

A second Multi-camera module (GDX_IS1_LR), with linear voltage regulators but maintaining the same architecture, was developed to reduce the high noise levels observed in the first module. Replacing the voltage regulators highly reduced the observed linear pattern and its frequency peak (Fig 16 (right), green curve). This improved the resulting SNR (Fig 16 (left), green), leading to a similar SNR levels to the reference commercial camera (Fig 16 (left), blue).

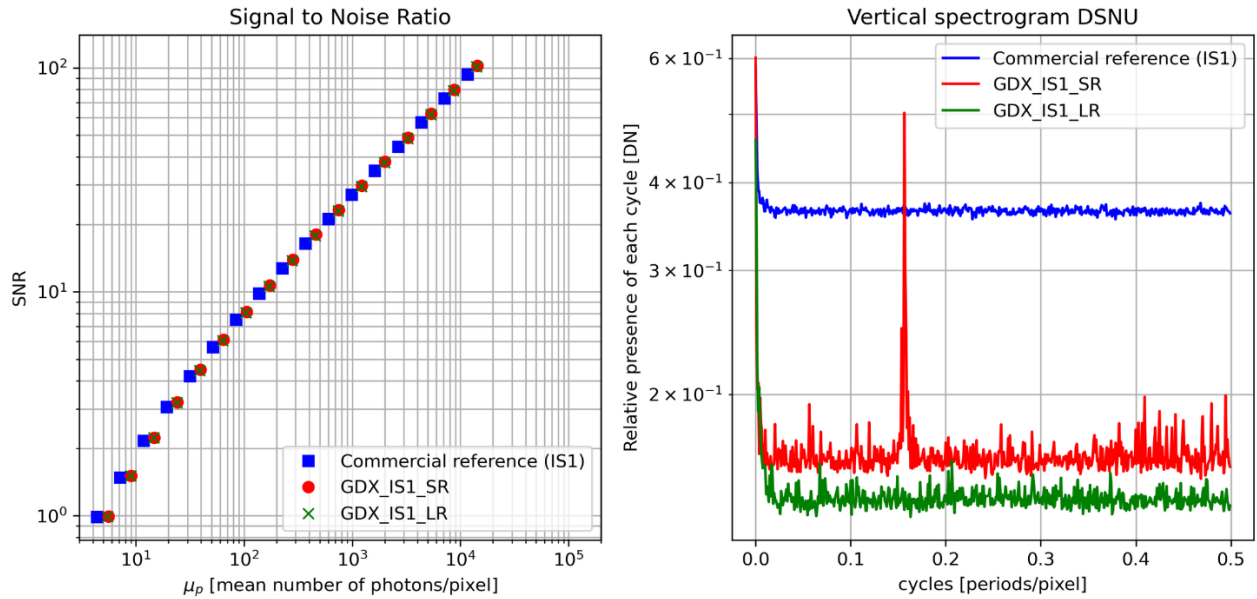

**Fig 16.** Signal to Noise Ratio (SNR) (Left) and Dark Signal Non-Uniformity (DSNU) (Right) plots from the developed GDX_IS1_LR (Red) and GDX_IS1_SR (Green) multi-camera modules and the reference commercial camera with IS1 sensor (Blue).

## Temperature and humidity influence

To determine temperature and humidity influence on the developed multi-camera module, the visible light characterization was repeated with the detector inside a climatic chamber under multiple controlled temperature and relative humidity conditions (the whole test setup including light source and integrating sphere was also mounted inside the climatic chamber). The measurements were done at 25°C (<20% RH), 50°C (<20% RH), 50°C (>90% RH) and compared to the measurement results at room conditions. In order to maintain the exact environmental conditions during all the stages of the light characterization measurements, it was decided not to open the chamber to place the sensor cover for the dark images measurement. Thus, the dark images measurements were done under the climatic chamber darkness conditions.

The results show no significant degradation of the signal to noise ratio (**Fig 17** (left)). The vertical spectrogram of dark signal non-uniformity (**Fig 17** (right)) exhibits no specific artifacts under the variation of the climatic conditions. Curves corresponding to measurement into climatic chamber (**Fig 17** (right), Blue, Red, Cyan) have an offset compared to the reference measurement at room conditions (**Fig 17** (right), Green). This was caused by the imperfect darkness conditions inside the climatic chamber.

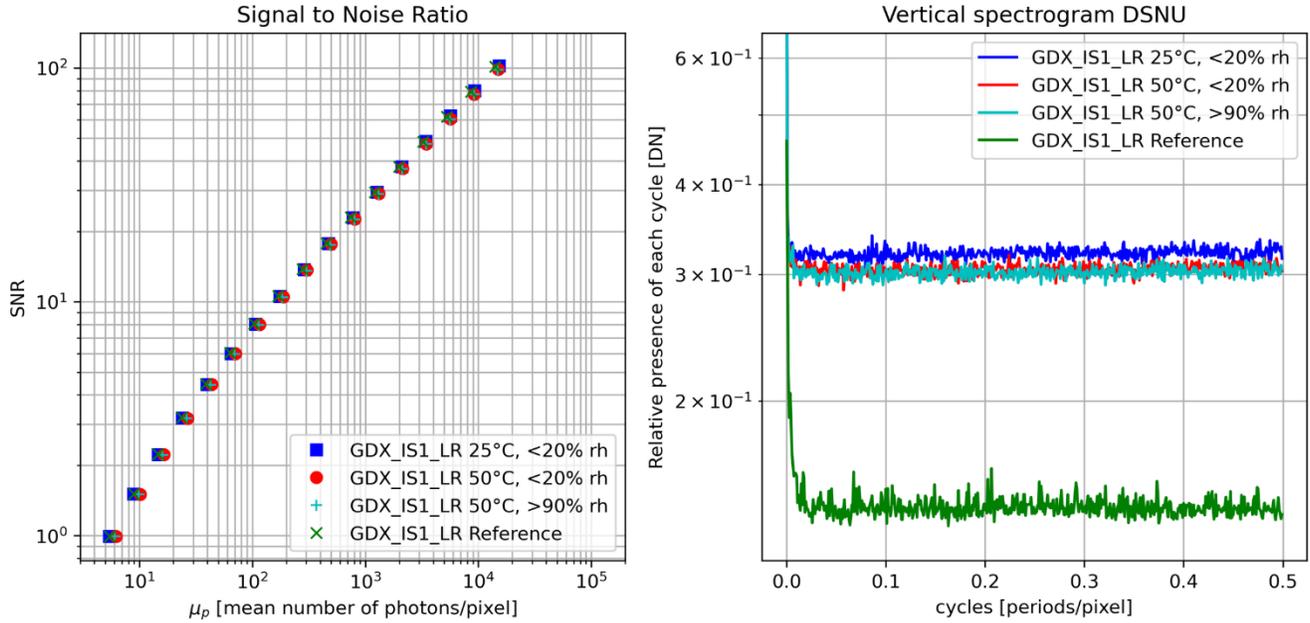

**Fig 17.** Signal to Noise Ratio (SNR) (Left) and Dark Signal Non-Uniformity (DSNU) (Right) plots from the developed GDX_IS1 multi-camera modules at room conditions (Green), 25°C and <20% RH (Blue), 50°C and <20% RH (Red), 25°C and >90% RH (Cyan).

## X-ray characterization

The developed detector was characterized with the two most promising scintillators identified during the characterization phase, i.e. the SC1(CI) (CsI:Ti) and the SC5(GOS) ($Gd_2O_2S$:Tb). The DQE measured for different DAK between 0.59 and 18.7 µGy peak at a frequency 0.5 $mm^{-1}$ (**Fig 18**). The low-frequency fixed pattern noise on the images increased the NPS and decreased the DQE below 0.5 $mm^{-1}$. The maximal DQE for the SC1(CI) and SC5(GOS) scintillators were around 0.60 and 0.30, respectively, for the reference DAK chosen at 2.34 µGy. The fixed pattern noise increases with the DAK squared and make the low-frequency DQE decreases as a function of the DAK.

The higher DQE obtained with the SC1(CI) scintillator can be explained by its micro-pillar structure and high thickness: CsI:Ti is a crystal that can be grown in vertical micro-pillars structures (2-5 um diameter, and over 400 um long) on a plane surface. This helps guiding the light in one direction and avoid lateral scattering, therefore, the MTF is generally better in CsI:Ti scintillators. $Gd_2O^2S:Tb$ is deposited in a powder structure and therefore has no light guiding capability, which leads to wide-angle scattering of the visible photons and lower image sharpness [27].

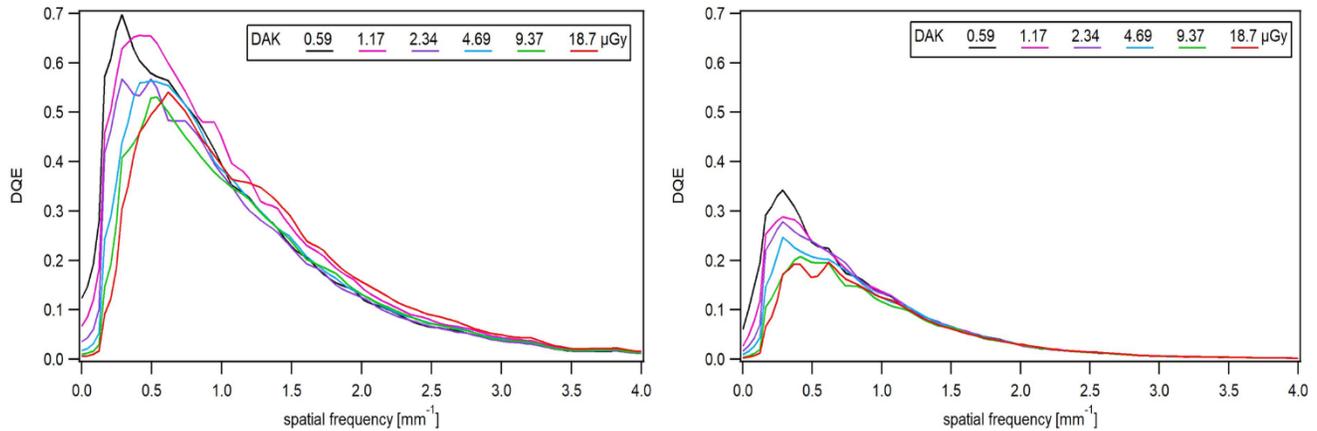

**Fig 18.** DQE curves obtained with the developed detector (GDX_IS1_LR) at different DAK. The left plots were obtained with the SC1(CI) Scintillator and the ones in the right with the SC5(GOS) scintillator.

The light throughput of the scintillator depends on the conversion efficiency for a given X-ray radiation. It is linked to the absorption rate and light yield (also called gain). For a given thickness, CsI:Ti has higher absorption rate and light yield compared to $Gd_2O^2S:Tb$. Therefore, the light throughput of CsI:Ti scintillators is higher [28]. Nevertheless, these performances have a cost. Due to its manufacturing complexity (crystal growth) that requires complex and expensive facilities, CsI:Ti scintillators are more expensive than $Gd_2O^2S:Tb$, especially for large area (e.g. 430 mm by 430 mm). Moreover, contrary to CSI:Ti, $Gd_2O^2S:Tb$ has very good chemical durability, mechanical properties, uniformity and is easier to manufacture [29].

In sum, compared to the SC1(CI) scintillator, the lower x-ray capture efficiency and x-rays to light photons conversion rate of the SC5(GOS) scintillator decrease the low-frequency DQE. The CsI needles of the SC1(CI) scintillator channel the light photons and avoid light spread that occurs in the grain structure of the SC5(GOS). Light spread decreases the spatial resolution and the high-frequency DQE of the SC5(GOS) scintillator.

# CONCLUSIONS

A novel, robust and low cost X-ray imaging system was developed, adapted to the needs and constraints of low and middle-income countries. The developed system is based on an indirect conversion chain: a scintillator plate produces visible light when excited by the X-rays then a matrix of the developed multi-camera modules converts the visible light from the scintillator into a set of digital images. The partial images are then unwarped, enhanced and stitched by a specialized software running on a network of FPGAs controlled by a master unit. By implementing a network of FPGA units (instead of a single more powerful processing unit), the partial images can be processed in parallel, reducing the total processing time and increasing the detectors frame-rate. Different commercially available components were characterized and the most suitable were implemented in the fabrication of the detector. The use of "off the shelf" components has the advantage of state of the art technologies at a lower cost than customized components (due to mass production) and the possibility of easily acquiring spare parts around the world.

The developed system was characterized at the Institute of Radiation Physics (IRA) from the Lausanne University Hospital (CHUV) using different standard medical radiology measurement setups. The characterization measurements of the developed detector led to high quality medical diagnostic images with DQE levels up to 60 % (@ 2.34 μGy).

Amongst the advantages of the developed detector are: Robustness: The detector was designed to withstand the hash environmental conditions of tropical countries. The characterization results at high temperature (50ºC) and humidity (>90% RH) conditions did not show any significant degradation of the signal to noise ratio. Low cost: the system fabrication costs are between 20 to 50% lower than the flat panel solutions available in the market today. Additionally, it is expected that, if mass-produced, the costs will decrease further. Modularity: The implemented modular design allows the system to be easily repaired (by simple replacement of individual modules) without the need of high technical expertise. Additionally, by using "off the shelf" components, the replacement parts can be easily obtained in the market and be replaced at a low costs, i.e. in the range of hundreds of USD. Conversely, when the detector from a traditional digital X-ray device breaks it must be replaced completely by specialized personnel at a high costs, i.e. in the range of tens of thousands USD.

However, these advantages come with a cost: due to the lower optical coupling of the implemented architecture, the DQE levels of the developed detector are around 15% lower than the flat panel solutions, i.e. the average DQE levels for commercial flat panels are around 70% (CsI:Ti) and 35% ($Gd_2O_2S$:Tb) while the DQE levels measured with our detector are 60% (CsI:Ti) and 30% ($Gd_2O_2S:Tb$). Nevertheless, the measured DQE levels are much higher than other multi-camera systems available in the market (The DQE levels for characterized commercial multi-camera detectors with CsI:Ti are between 30 to 40%). Therefore, the developed X-ray detector shows very promising results and potential for being implemented in the context and harsh environmental conditions of low and middle-income countries at a lower purchasing and maintenance costs than traditional digital X-ray detectors.

## ACKNOWLEDGMENTS

The authors would like express their gratitude to the Swiss Innovation Agency (Innosuisse) for the financial support of this project.